# Fundamental thickness limit of itinerant ferromagnetic SrRuO$_3$ thin films

Young Jun Chang[1], Choong H. Kim[2], S.-H. Phark[1], Y. S. Kim[1], J. Yu[2], and T. W. Noh[1*]

[1] *ReCOE & FPRD, Department of Physics and Astronomy, Seoul National University, Seoul 151-747, Korea.*

[2] *CSCMR & FPRD, Department of Physics and Astronomy, Seoul National University, Seoul 151-747, Korea*

*e-mail: twnoh@snu.ac.kr

We report on a fundamental thickness limit of the itinerant ferromagnetic oxide SrRuO$_3$ that might arise from the orbital-selective quantum confinement effects. Experimentally, SrRuO$_3$ films remain metallic even for a thickness of 2 unit cells (uc), but the Curie temperature, $T_C$, starts to decrease at 4 uc and becomes zero at 2 uc. Using the Stoner model, we attributed the $T_C$ decrease to a decrease in the density of states ($N_o$). Namely, in the thin film geometry, the hybridized Ru-$d_{yz,zx}$ orbitals are terminated by top and bottom interfaces, resulting in quantum confinement and reduction of $N_o$.

SrRuO$_3$ is an itinerant ferromagnetic oxide with a ferromagnetic Curie temperature ($T_C$) of 160 K and a magnetic moment of 1.6 μ$_B$ [1,2]. At $T_C$, its electrical resistivity curve changes slope due to the decrease in spin fluctuations [3]. The ferromagnetic metallic ground state can be understood in the framework of the Stoner model [3], which is based on the competition between the kinetic energy and exchange energy due to Coulomb repulsion. The stabilization of a ferromagnetic state is favored when

$$I\,N_0 > 1 \, , \qquad (1)$$

where $N_0$ and $I$ are the non-magnetic density-of-state (DOS) at the Fermi level $E_F$ and the effective electron-electron interaction, respectively. The model also predicts that

$$T_C \propto [1 - 1/(I\,N_0)]^{1/2} \, . \qquad (2)$$

Note that a systematic variation of $N_0$ can induce $T_C$ change in the band ferromagnetic material.

In simple metals, ultrathin films are known to exhibit quantum confinement (QC) effects due to the confinement of electrons inside the conducting layers [4]. These effects alter the electronic structure and $N_0$, so they can result in striking variations of the physical properties such as conductivity, magnetism, and the Hall effect. However, up to this point, no QC effects have been reported in oxide ultrathin films, including SrRuO$_3$.

SrRuO$_3$ is also one of the most frequently used oxide electrode materials. It has high chemical stability, good thermal properties [5], and the perovskite crystal structure, which are advantageous for integration with other oxide materials for the fabrication of heterostructures [6]. SrRuO$_3$ has been used in numerous oxide applications, including Josephson junctions [7], spin-polarized ferromagnetic tunneling junctions [8], Schottky junctions [9], field effect devices [10], ferroelectric capacitors [11,12], and multiferroic devices [13]. To design such oxide thin film devices, it is important to understand how thickness ($t$) affects the transport and magnetic properties of SrRuO$_3$, especially in the ultrathin limit.

In this Letter, we report on the thickness-dependent ferromagnetism of ultrathin SrRuO$_3$ films. We observed that the metallic properties of SrRuO$_3$ ultrathin films are maintained down to $t$ = 2 uc, but $T_C$ starts to decrease at $t$ = 4 uc and becomes zero at $t$ = 2 uc. To explain the $T_C$ reduction, we investigated the variation of $N_0$ using *in situ* scanning tunneling spectroscopy (STS) and the first principles calculations. Both of these experimental and theoretical studies showed that there is a significant decrease in the DOS at $E_F$ for $t$ < 4 uc. This behavior could be explained in terms of QC of Ru-$d_{yz,zx}$ orbitals.

To obtain high-quality SrRuO$_3$ ultrathin films, we used a TiO$_2$-terminated SrTiO$_3$

(001) substrate, the surface of which is atomically flat with 1-uc-high steps (small miscut, i.e., <0.1°). On top of the substrate, we deposited SrRuO$_3$ films using pulsed laser deposition (PLD) at 700°C with an oxygen pressure of 0.1 Torr and a laser fluence of 2.5 J/cm$^2$ [14]. The growth rate of SrRuO$_3$ films was about 0.015 nm/sec. The film thickness was controlled by *in situ* monitoring of the reflection high-energy electron diffraction intensity oscillation while a two-dimensional (2D) growth mode was maintained. Ambient atomic force microscopy (AFM) topography of the SrRuO$_3$ films shows atomically smooth surfaces that resemble the substrate surface, as shown in Fig. 1(a)-(c), indicating the step-flow growth mode [15,16].

Figure 1(e) shows the temperature-dependent dc resistivity $\rho(T)$ for SrRuO$_3$ films for various values of $t$. As $t$ decreases, the $\rho(T)$ curves show systematic changes: namely, the $\rho$(20 K) value becomes higher and the slope change occurs at a lower temperature. It should be noted that our SrRuO$_3$ films remain metallic down to $t = 2$ uc. As stated earlier, $\rho(T)$ should show an anomaly at $T_C$. As indicated by the arrows in Fig. 1(e), the $\rho(T)/\rho$(300 K) data also show anomalies, which we attribute to the ferromagnetic ordering. We estimated the $T_C$ values by differentiating the $\rho(T)/\rho$(300 K) curves. The black solid circles in Fig. 2(a) show the experimental $T_C$ data for the SrRuO$_3$ films. Above 4 uc, $T_C$ remains nearly constant. However, $T_C$ decreases drastically below 4 uc, and it becomes zero for the 2-uc film.

There have been earlier experimental reports that a metal-insulator transition should occur at either 3 or 4 uc of SrRuO$_3$ film [17,18]. Note that our observation on the persistence of the metallic state down to 2 uc is not consistent with the earlier experimental reports but agrees with the first-principles calculations [19]. To obtain further insight, we deposited SrRuO$_3$ ultrathin films at a higher growth rate, 0.045 nm/sec. As shown in Fig. 1(d), the surface morphology of the films is very rough, indicating a 3D island growth mode even for the film with $t$ = 4 uc. As shown in the inset of Fig. 1(e), the rough film has an insulating behavior. Therefore, the difference between our work and earlier work could be attributed to the effects of disorder, such as grain boundaries and step edges.

Using our high quality SrRuO$_3$ thin films, we investigated the thickness dependence of the low temperature resistivity, i.e., $\rho$(20 K). Our 46-uc film has $\rho$(20 K) = 51 μΩcm, which is comparable to its bulk value of 20 μΩcm [1]. As shown in Fig. 2(b), $\rho$(20 K) increases systematically with a decrease in $t$. According to the Drude model for a metal, $\rho = m^*/ne^2\tau$, where $m^*$, $n$, and $\tau$ are the effective mass, number density, and scattering time of the free carriers, respectively. Down to 4 uc, the increase in $\rho$(20 K) can be explained in terms of the enhanced surface scattering rates in a film geometry. Namely, the effect of $t$ can be incorporated by including additional channel of scattering at the top and the bottom surfaces of the thin film. According to Matthiessen's rule, the net $\rho$ is described by a combined $\tau$ with contributions from the bulk and the surface scattering [20]. Assuming that the surface scattering determines the minimum value of electron collision time in the thin film, we have

$$\rho = \rho_b + \left(\frac{m^* v_F}{ne^2}\right)\frac{1}{t}, \qquad (3)$$

where $\rho_b$ and $v_F$ are the bulk resistivity and the Fermi velocity, respectively. Using the reported bulk values of $\rho_b(20\ \text{K}) \approx 20\ \mu\Omega\text{cm}$, $m^* \approx 7 m_e$, $v_F \approx 2 \times 10^7$ cm/s, $n \approx 1.2 \times 10^{22}$ /cm$^3$, and $\tau \approx 5 \times 10^{-15}$ sec [1,21,22], Eq. (3) describes the $t$-dependent $\rho(20\ \text{K})$ for our SrRuO$_3$ films with $t > 4$ uc. The solid line in Fig. 2(b) is the classical theoretical prediction based on the increased surface scattering. For the films with $t = 2$ and 3 uc, $\rho(20\ \text{K})$ becomes even higher than the classical predictions. A comparison with Eq. (3) indicates the possibility of decreasing $n$ for $t < 4$ uc.

To obtain experimental evidence of the $t$-dependent change of $N_0$, we carried out *in situ* STS studies on 1- to 6-uc-thick SrRuO$_3$ films, which were grown on semiconducting Nb(0.1%)-doped SrTiO$_3$ substrates. Figure 3 shows the tunneling current ($I_t$) versus tip bias voltage ($V_{tip}$) at room temperature, obtained in an ultra-high vacuum scanning tunneling microscopy chamber connected in a vacuum to the PLD system. The red line corresponds to the $I_t$-$V_{tip}$ spectra of a 50-uc-thick film, which is expected to show nearly the same metallic behavior as the bulk material. As $t$ increases, the corresponding $I_t$ becomes larger at a given $V_{tip}$. The inset of Fig. 3 shows the tunneling conductance ($dI_t/dV_{tip}$) values at zero bias, which should be proportional to $N_0$ [23]. As $t$ decreases, the $dI_t/dV_{tip}$(0 V) values decrease monotonically. In particular, the decrease becomes quite significant below 4 uc. In addition, $dI_t/dV_{tip}$(0 V) remains finite for the 1-uc film. Using the Stoner theory in Eq. (2), the $T_C$ values were evaluated and plotted as blue triangles in Fig. 2(a). The excellent agreement with the $T_C$ values from the $\rho(T)$ curves indicates that the decrease in $N_0$ in SrRuO$_3$ ultrathin films should play an important role in their ferromagnetic properties [24].

To understand the thickness dependence of $N_0$, we carried out first-principles calculations based on the density functional theory. We carried out first-principles calculations for SrRuO$_3$ thin films on a SrTiO$_3$ substrate using the Vienna *Ab initio* Simulation Package (VASP) with the Ceperley-Alder parameterization of the local spin density approximation and projector-augmented wave potentials. The cutoff energy for the plane waves was 500 eV. We sampled the *k*-points from the 12×12×1 Monkhorst-Pack (MP) grid and performed integrations with 0.2 eV Gaussian smearing. To calculate more accurate pDOS, we also used the 24×24×1 MP *k*-point grid and tetrahedron method. In our calculations, we did not consider the effect of on-site Coulomb interactions. We used 3.5-uc $SrTiO_3$ (4 SrO layers and 3 TiOs layers) slabs sandwiched symmetrically by 1- to 6-uc $SrRuO_3$ slabs. We fixed the in-plane and out-of-plane lattice constants of the ultrathin films to the experimental values, $a$ = 0.3905 nm and $c$ = 0.3959 nm, respectively.

Figure 4(a) displays schematic pictures of the samples used for the calculations: 1- and 4-uc $SrRuO_3$ films on 2-uc $SrTiO_3$. Initially, we calculated the projected density-of-state (pDOS) for the non-magnetic ground states of each $SrRuO_3$ layer in the $SrRuO_3$ films with $t$ = 1 to 4 uc. The four gray plots in the middle of Fig. 4(b) show the non-magnetic pDOS projected onto a layer close to the center of the $SrRuO_3$ thin film. The 1-uc film shows a few sharp peaks around $E_F$, i.e., near 0 eV. As the film becomes thicker, a larger number of peaks remain, and the strongest peak moves toward $E_F$. In addition, the DOS at $E_F$, i.e., $N_0$, becomes larger. Using these DOS values, the $T_C$ values were also estimated with Eq. (2), and these are plotted as red squares in Fig. 2(a).

The systematic variation of the DOS at $E_F$ in the non-magnetic states can be understood in terms of the orbital-selective QC effects. Note that the $Ru^{4+}$ ion has four electrons in the $t_{2g}$ orbitals, which are extended in the $xy$-, $yz$-, or $zx$-planes. Due to the anisotropic shape of the $t_{2g}$ orbitals, they can orient in a particular direction and provide directional hybridizations with surrounding oxygen ions. In bulk SrRuO$_3$, all the $t_{2g}$ orbitals form 2D networks of individual $d_{xy}$, $d_{yz}$, and $d_{zx}$ orbitals by hybridizing with the oxygen 2$p$ orbitals in each plane and form 2D tight-binding bands [25]. Figure 4(c) shows the schematic view of the Ru $t_{2g}$ orbitals for the 1-uc sample. The (red) $d_{xy}$ orbitals are extended in the $xy$-plane, so the corresponding pDOS should have a 2D van Hove singularity, shown as the red-shaded area at the top of Fig. 4(b). On the other hand, the (blue) $d_{yz}$ (or $d_{zx}$) orbitals should be extended in the $yz$- ($zx$-) plane, but they are truncated by the top and bottom interfaces. This QC effect in the $d_{yz}$ and $d_{zx}$ orbitals will result in a pDOS with strong 1D singularities near the band edges, shown as the blue-shaded area at the top of Fig. 4(b). Note that the pDOS for the 1-uc SrRuO$_3$ film determined from first-principles calculations can be explained as the sum of the pDOS for the $d_{xy}$, $d_{yz}$, and $d_{zx}$ orbitals. A similar argument can be made for the 4-uc sample, as shown in Fig. 4(d). In this case, the resulting pDOS of the $d_{yz}$ and $d_{zx}$ orbitals should have eight peaks, shown in the blue shaded area at the bottom of Fig. 4(b). The success of these simple arguments suggests that the calculated changes in the pDOS shown in Fig. 4(b) can be interpreted in terms of the orbital-selective QC effects.

The systematic change in pDOS of the SrRuO$_3$ films should also affect their magnetic properties. We calculated the pDOS for the SrRuO$_3$ films in ferromagnetic configurations, as shown in Fig. 4(e). For the thicker films, i.e., $t = 4$ and 3 uc, the calculations show that there is a large asymmetry in the pDOS between the spin-up and spin-down states, supporting the Stoner-type ferromagnetic ground state. The calculated values of the magnetic moment are 0.42 and 0.14 $\mu_B$/Ru for $t = 4$ and 3 uc, respectively. On the other hand, for the thinner films, i.e., $t = 2$ and 1 uc, the non-magnetic states are favored energetically. These predictions explain how $T_C$ becomes zero for our SrRuO$_3$ sample with $t = 2$ uc.

In summary, we observed that the ferromagnetic Curie temperature of SrRuO$_3$ ultrathin films vanishes for the 2-unit-cell samples. Using *in situ* scanning tunneling spectroscopy and first-principles calculations, we showed that such a decrease might originate from a decrease in the density of states at the Fermi level, an effect due to orbital-selective quantum confinement effects. This work suggests that the properties of itinerant ferromagnetic materials can be controlled by geometric constriction at the nanoscale level.

**Acknowledgements**

This research was supported by Basic Science Research Program through the National Research Foundation of Korea (**NRF**) funded by the Ministry of Education, Science and values show a monotonic decrease for decreasing $t$ and remain finite down to 1 uc. The red solid line indicates the $dI_t/dV_{tip}(0\ V)$ value for a 50-uc film. Each curve is averaged over an area of 100 nm × 100 nm by a grid spectroscopic mode with 20 × 20 sampling pixels. All STS measurements were carried out at room temperature with the same tip under identical scanning conditions, namely, $I_t = 0.1$ nA and $V_{tip} = 0.5$ V.

**Figure legends**

FIG. 1: (color online). (a)-(c) AFM topography of $SrRuO_3$ thin films for different thickness values ($t$) of 2 (a), 4 (b), and 27 (c) unit cells (uc). (d) Topography of rough film with $t$ = 4 uc obtained from much faster growth rate. All images are 2 μm × 2 μm in size. (e) Temperature ($T$) dependence of the normalized resistivity $\rho(T)/\rho(300\,K)$ of $SrRuO_3$ thin films for $t$ of 2 to 46 uc. The $SrRuO_3$ films remain metallic down to 2 uc. The inset shows the $\rho(T)/\rho(300\,K)$ of the rough film with $t$ = 4 uc.

FIG. 2: (color online). (a) $t$ dependence of $T_C$ values. Experimental $T_C$ values (black solid circles) were obtained from the peak $T$ positions in the inset of Fig. 1(b). $T_C$ values were also estimated based on the Stoner model, i.e., Eq. (2), using values for the non-magnetic density-of-state at the Fermi level ($N_0$) estimated from *in situ* STS studies (blue open triangles) and first-principles calculations (red open squares). (b) $t$-dependence of $\rho(20\,K)$. The solid line is the classical theoretical prediction based on the increased surface scattering.

FIG. 3: (color online). Plot of the tunneling current $I_t$ versus tip bias voltage $V_{tip}$ obtained from $SrRuO_3$ films grown on a Nb-doped $SrTiO_3$ substrate. The inset shows the $t$ dependence of the tunneling conductance ($dI_t/dV_{tip}$) at $V_{tip}$ = 0 V. The $dI_t/dV_{tip}(0\,V)$

FIG. 4: (color online). (a) Schematic pictures of the model systems: 1 and 4 uc of SrRuO$_3$ on 2 uc of SrTiO$_3$. The green and blue squares represent the metal-oxygen octahedra with Ru$^{4+}$ and Ti$^{4+}$ ions, respectively. The symbol of S-$i$ indicates the $i^{th}$ SrRuO$_3$ layer from the surface. (b) Calculated non-magnetic projected density-of-state (pDOS) (gray) for the middle SrRuO$_3$ layer in films with $t = 1 - 4$ uc. Note that the pDOS at 0 eV decreases with decreasing $t$. (c) Orbital configurations of the $t_{2g}$ electrons of Ru$^{4+}$ ions in the 1 uc SrRuO$_3$ film. The $d_{xy}$ orbitals (red) are hybridized with the oxygen 2$p$ orbitals (not shown) and form an infinite 2D sheet. On the other hand, because of the truncation in the $z$-direction, the $d_{yz}$ or $d_{zx}$ orbitals (blue) form a 1D strip. The Ti-oxygen octahedra are shown in sky blue. (d) Orbital configurations of the $t_{2g}$ electrons of Ru$^{4+}$ ions in the 4-uc SrRuO$_3$ film. (e) Calculated pDOS with ferromagnetic configurations for the middle SrRuO$_3$ layer in films with $t = 1–4$ uc. The majority and minority spin states are colored in blue (filled) and red (unfilled), respectively.

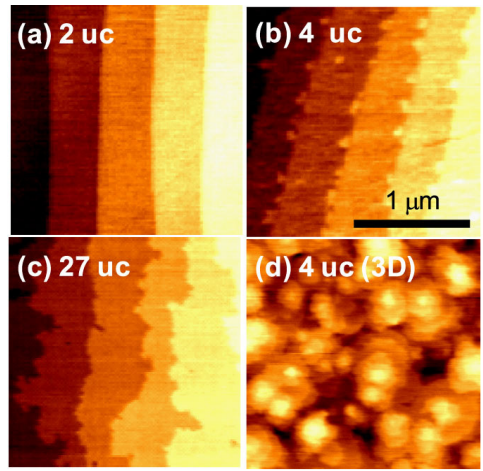

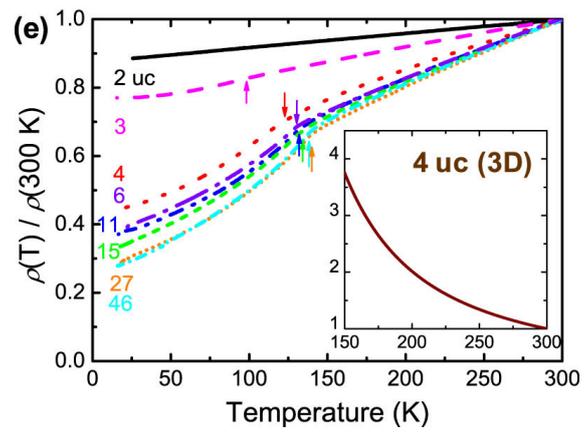

**Chang *et al*., Fig. 1**

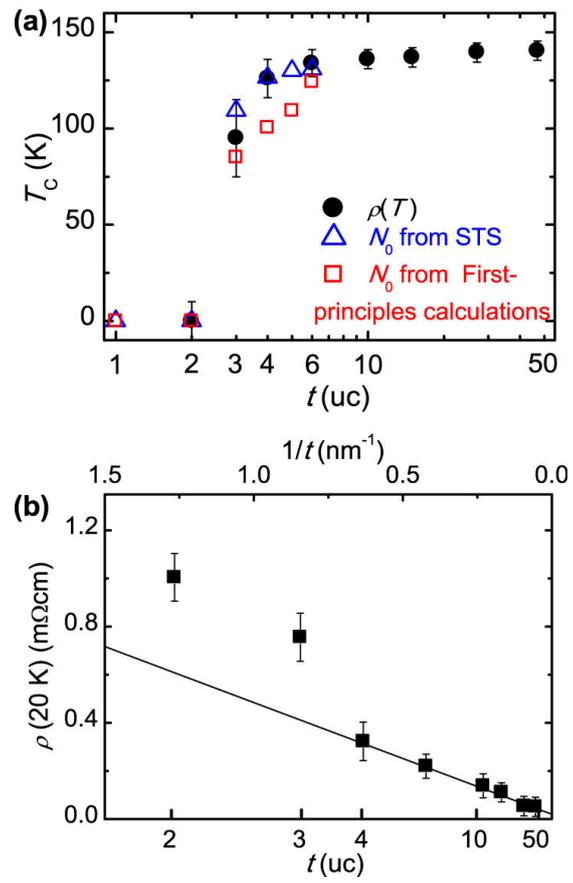

**Chang *et al*., Fig. 2**

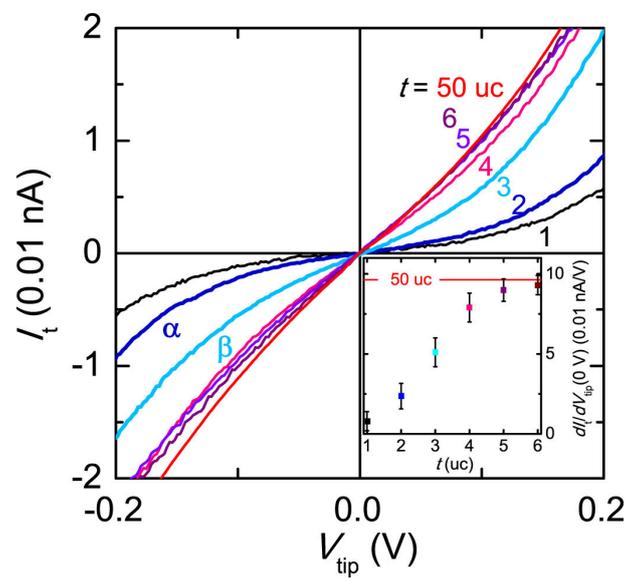

**Chang *et al*., Fig. 3**

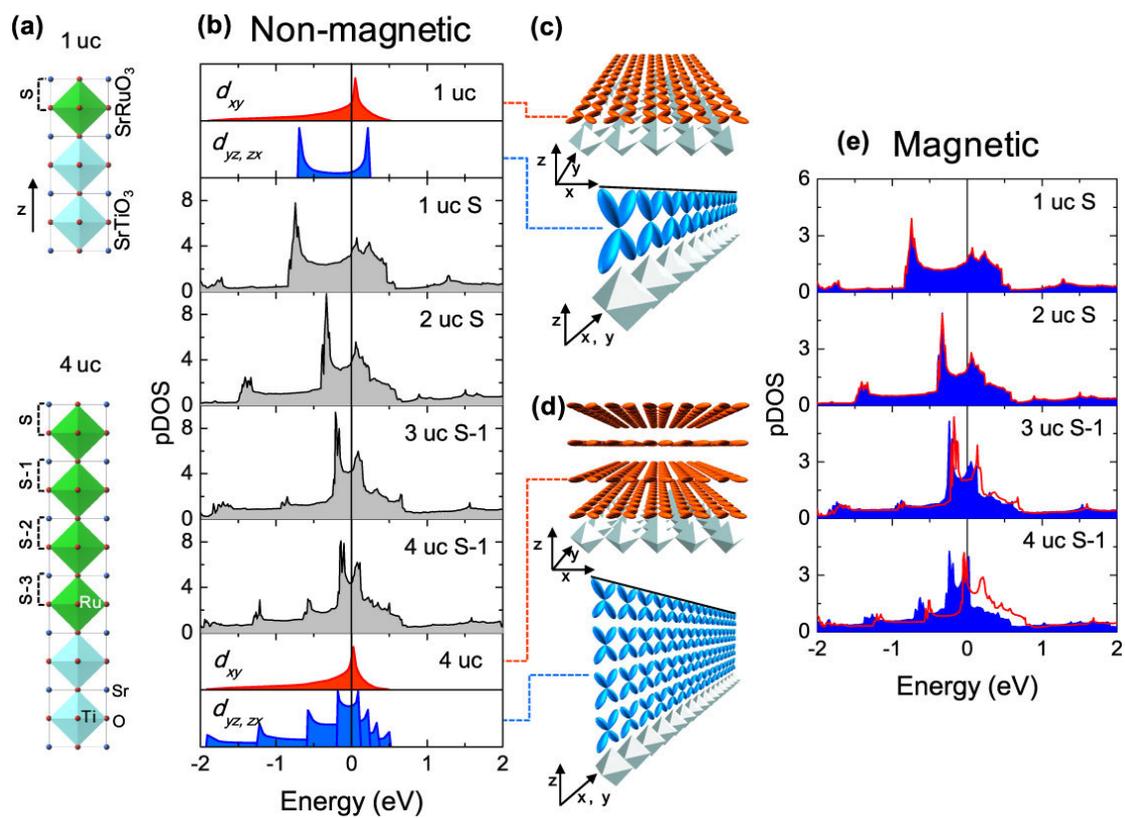

**Chang *et al*., Fig. 4**